\documentclass[aps,pre,superscriptaddress,twocolumn,amsmath,amssymb,showpacs]{revtex4}
\usepackage{graphicx}
\usepackage{dcolumn}
\usepackage{bm}

\begin{document}

\title{Fractional Fokker-Planck dynamics: Numerical algorithm and simulations}

\author{E. Heinsalu}
  \affiliation{Institut f\"ur Physik,
  Universit\"at Augsburg,
  Universit\"atsstr. 1,
  D-86135 Augsburg, Germany}
  \affiliation{Institute of Theoretical Physics, Tartu University,
  T\"ahe 4, 51010 Tartu, Estonia}

\author{M. Patriarca}
  \affiliation{Institut f\"ur Physik,
  Universit\"at Augsburg,
  Universit\"atsstr. 1,
  D-86135 Augsburg, Germany}

\author{I. Goychuk}
  \affiliation{Institut f\"ur Physik,
  Universit\"at Augsburg,
  Universit\"atsstr. 1,
  D-86135 Augsburg, Germany}

\author{G. Schmid }
  \affiliation{Institut f\"ur Physik,
  Universit\"at Augsburg,
  Universit\"atsstr. 1,
  D-86135 Augsburg, Germany}

\author{P. H\"anggi}
  \affiliation{Institut f\"ur Physik,
  Universit\"at Augsburg,
  Universit\"atsstr. 1,
  D-86135 Augsburg, Germany}

\date{\today}

\begin{abstract}
Anomalous transport in a tilted periodic potential is investigated
numerically within the framework of the fractional Fokker-Planck
dynamics via the underlying CTRW. An efficient numerical algorithm is
developed which is applicable for an arbitrary potential.
This algorithm is then applied to investigate the fractional current and  the corresponding nonlinear mobility in
different washboard potentials.
Normal and fractional diffusion are compared through their time evolution
of the probability density in state space. 
Moreover, we discuss the stationary probability density of the fractional current values.
\end{abstract}

\pacs{ 05.40.-a, 02.70.-c, , 02.50.Ey}

\maketitle


\section{Introduction}

Thermal diffusion of Brownian particles under the action of a
periodic force continues to present an active field of research over recent
years, being relevant for various applications in condensed matter
physics, chemical physics, nanotechnology, and molecular biology
\cite{100years,HAM,HM,freund,sokolov,HTB}.

The stochastic motion of Brownian particles in a potential
\begin{eqnarray} \label{totpot}
U(x) = V (x) - Fx \, ,
\end{eqnarray}
where $V (x) = V (x+L)$ is the periodic substrate potential with
period $L$ and $F$ is the constant bias, is qualitatively well
known \cite{HTB,BM}: Particles, subject to friction and noise,
will diffuse and drift in the direction of the applied bias. In the overdamped
regime and in the absence of noise, the particles perform a
creeping motion. If the tilting force $F$ is large enough, so that
the total potential $U(x)$ has no minima, the particles move down
the corrugated plane. If minima do exist, the particles arrive there and stop.
In the presence of noise, the particles do not stay
permanently in the minima but will undergo noise activated escape events. The
particles thus perform a hopping process from one well to the neighboring ones.

It is well know that among many other applications \cite{HTB,BM}, the model of a
Brownian particle in a periodic potential can be used to describe
Brownian motors and molecular motors \cite{astumian,reimann,nori,julicher}, such as kinesins, dyneins, and
myosins. However, many other systems, such as RNA polymerases,
exonuclease and DNA polymerases, helicases, the motion of
ribosomes along mRNA, the translocation of RNA or DNA through a
pore, are advantageously  described as particles moving along a disordered
substrate. Depending on the statistical properties of the
potential, the long-time limit of the process can be
quite different from that in a washboard potential
\cite{nelson2004}. It has been shown that the heterogeneity of the
substrate potential may lead to anomalous dynamics \cite{haus87,bouchaud90a,bouchaud90b, derrida1983}.
In particular, over a range of forces around the stall force subdiffusion
is observed \cite{nelson2004}, i.e. the displacement grows as
$\langle \delta r^2 (t) \rangle \sim t^\alpha$, with $0 < \alpha < 1$.

Given the importance of the subdiffusion and the motion in
periodic potentials in various applications \cite{reimann,sokolov05,metzler2000}, and considering the
biological systems mentioned above, we address the physics  of the
effect of the combined action of a biased periodic force and a
random substrate. Within our approach  we model
the subdiffusive dynamics in terms of a suitable residence time probability
density with a long tail \cite{bouchaud90a,bouchaud90b, scher75} rather than through a random potential
\cite{haus87,bouchaud90b,schimansky}.


The  Fokker-Planck equation describing the overdamped
Brownian motion in the potential $U(x)$ can be generalized to
anomalous transport.  The corresponding result is known as the {\it fractional
Fokker-Planck equation}  \cite{metzler2000, metzler99, barkai2001},
being the central equation of fractional dynamics,
\begin{eqnarray}\label{FFPERL}
\frac{\partial}{\partial t} P(x,t) = \sideset{_0}{_t}{\mathop{\hat
D}^{1-\alpha}} \left [ \frac{\partial}{\partial x}
\frac{U'(x)}{\eta _\alpha} + \kappa _\alpha \frac{\partial ^2}{\partial x^2}
\right ] P(x, t) \, .
\end{eqnarray}
In our notation $P(x, t)$ is the probability density, a prime stands for the
derivative with respect to the space coordinate, $\kappa _\alpha$ denotes
the anomalous diffusion coefficient with physical dimension
$[\mathrm{m}^2 \mathrm{s}^{-\alpha}]$. The quantity $\eta_\alpha$ denotes the
generalized friction coefficient possessing the dimension
$[\mathrm{kg} \, \mathrm{s}^{\alpha -2}]$; it is related to $\kappa _\alpha$
through $\eta_\alpha \kappa _\alpha = k_\mathrm{B} T$, thus constituting a generalized Einstein relation.

The notation $\sideset{_0}{_t}{\mathop{\hat D}^{1-\alpha}}$ on the right-hand
side of Eq.~(\ref{FFPERL}) stands for the integro-differential
operator of the Riemann-Liouville fractional derivative, defined
 as follows  \cite{metzler2000, Sokolov, GorenfloMainardi},
\begin{equation} \label{RL}
\sideset{_0}{_t}{\mathop{\hat D}^{1-\alpha}} P (x, t) =\frac{1}{
\Gamma(\alpha)} \frac{\partial}{\partial t} \int_{0}^{t}
\mathrm{d} t' \, \frac{P(x, t')}{(t-t')^{1-\alpha}} \, ,
\end{equation}
for $0 < \alpha < 1$. The Riemann-Liouville operator (\ref{RL})
introduces a convolution integral with a slowly decaying power-law
kernel, which is typical for memory effects in complex systems.
Equation~(\ref{FFPERL}) describes subdiffusive processes for $0 <
\alpha < 1$, and reduces to the ordinary Fokker-Planck equation
when $\alpha = 1$.

The fractional Fokker-Planck equation was originally introduced
with the Riemann-Liouville fractional derivative on its right-hand
side \cite{metzler2000, Sokolov}. However, it can sometimes be
more convenient to switch to an equivalent representation that involves the Caputo fractional
derivative. This formulation can provide genuine technical advantages. The
fractional Fokker-Planck equation can then be rewritten as:
\begin{eqnarray}\label{FFPE}
D_{*}^{\alpha} P(x,t)= \left [ \frac{\partial}{\partial x}
\frac{U'(x)}{\eta_\alpha} + \kappa _\alpha \frac{\partial
^2}{\partial x^2} \right ] P(x, t) \, ,
\end{eqnarray}
with the Caputo fractional derivative $D_{*}^{\alpha}$ on its
left-hand side \cite{GorenfloMainardi}; i.e.,
\begin{eqnarray}
D_{*}^{\alpha} P (x, t) = \frac{1}{\Gamma(1 - \alpha)} \int_0^t
\mathrm{d} t' \frac{1}{(t - t')^\alpha} \frac{\partial}{\partial
t'} P (x, t') \, .
\end{eqnarray}

In the present work we investigate the anomalous transport described by
the fractional Fokker-Planck equation (\ref{FFPERL}), (\ref{FFPE})
through the numerical simulation of the corresponding continuous time
random walk (CTRW).
In Sec.~\ref{sec-num} we develop the algorithm for the numerical simulations.
In Sec.~\ref{current} we study the fractional current and the mobility for various types of tilted periodic potentials,
and in Sec.~\ref{distribution} the time evolution of the probability density in space as well as the density of the current values.
In doing so we emphasize similarities and analogies between anomalous and normal diffusion.
The implications of the differences are discussed in the Conclusions.


\section{Numerical simulation of the fractional Fokker-Planck equation through the underlying continuous time
random walk} \label{sec-num}

The fractional Fokker-Planck equation represents the continuous
limit of a CTRW with the Mittag-Leffler
residence time density \cite{letter}
\begin{equation} \label{ML}
\psi_i(\tau) = -\frac{\mathrm{d}}{\mathrm{d} \tau} E_{\alpha}(
-(\nu_i \tau)^{\alpha}) \, ;
\end{equation}
$E_{\alpha}(-(\nu_i \tau)^{\alpha})$ is the Mittag-Leffler
function,
\begin{equation} \label{MLsurv}
E_{\alpha}(-(\nu_i \tau)^{\alpha}) = \sum_{n = 0}^{\infty}
\frac{[-(\nu_i \tau)^{\alpha}]^n}{ \Gamma(n \alpha + 1)} \, ,
\end{equation}
and the quantity $\nu_i^{-1}$ denotes the time-scaling parameter at
site $i$.

The numerical algorithm of the CTRW can be
readily implemented for the motion in an arbitrary force field, as
we will demonstrate below.


\subsection{Numerical algorithm for the continuous time
random walk} \label{sec-Alg}

\begin{figure}[t]
\centering
\includegraphics[width=6.5cm]{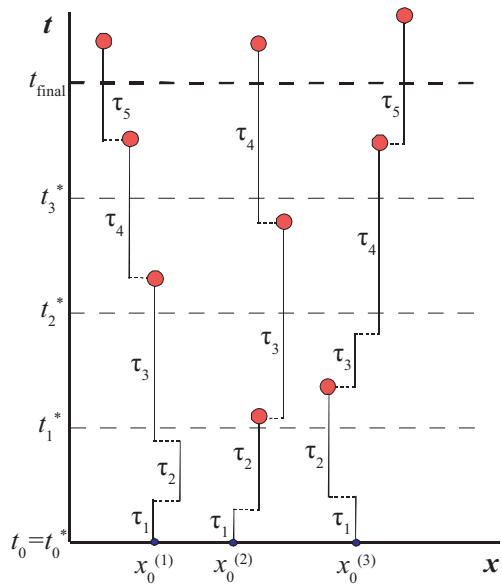}
\caption{(color online) Sketch of the numerical algorithm: After a random waiting time
$\tau$ the particle jumps from the current position $x^{(n)}$ to
the position $x^{(n)} + \Delta x$ or $x^{(n)} - \Delta x$. The
process is reiterated until $t^{(n)} \geq t_\mathrm{final}$. The
numerical measurements are performed after constant time intervals
at times $t_m^*$. The full-circles represent the events that are
used for the computation of the physical quantities (see text for an explanation). }
\label{algorithm}
\end{figure}

To study the CTRW in an one-dimensional
potential $U(x)$, we consider an ensemble of $N$ particles moving
on a lattice $\{x_i=i\Delta x \}$, with a lattice period $\Delta
x$; $i= 0, \pm 1, \pm 2, \ldots$ We emphasize that the numerical algorithm that
we provide is valid for an arbitrary potential, i.e. not only for the potential
(\ref{totpot}). The state of the $n$-th particle
is defined through its current position $x^{(n)}$ and the time
$t^{(n)}$ at which it will perform the next jump to a
nearest-neighbor site.

The $n$-th particle of the ensemble starts from the initial
position $x^{(n)}(t_0) = x_0^{(n)}$. After a residence time $\tau$
extracted from the probability density $\psi_i(\tau)$, the
particle jumps from site $i$ to site $i + 1$ or  $i - 1$ with
probability $q_i^{+}$ or $q_i^{-}$, respectively, obeying the
normalization condition $q_i^{+} + q_i^{-} =1$. Correspondingly,
the space coordinate and the time are updated, $x^{(n)} \to
x^{(n)} \pm \Delta x$ and $t^{(n)} \to t^{(n)} + \tau$.
Reiterating this procedure, the full random trajectory of the
random walker can be computed (see Fig.~\ref{algorithm}).

In order to perform the numerical simulation, one needs to
evaluate the quantities  $q_i ^{\pm}$ and $\nu _i$. In terms of the
fractional transition rates $f_i$ and $g_i$, they can be expressed
in the following way,
\begin{eqnarray}\label{q-nu}
q_i^+ &=& f_i/(f_i + g_i) \, , \qquad  q_i^- = g_i/(f_i + g_i)  \, , \\
\label{q-mu}
\nu_i &=& (f_i + g_i)^{1/\alpha} \,
,
\end{eqnarray}
where we have chosen
\begin{subequations} \label{f-rate}
\begin{align}
f_i &= (\kappa_{\alpha}/\Delta x ^2) \exp [- \beta(U_{i + 1} - U_i)/2] \, ,  \\
g_i &= (\kappa_{\alpha}/\Delta x ^2) \exp [- \beta (U_{i - 1} -
U_i)/2] \, .
\end{align}
\end{subequations}
Here $\beta = 1/k_\mathrm{B} T$ is the inverse temperature and
$U_i \equiv U(i \Delta x)$.
An appropriate discretisation step $\Delta x$ has to satisfy the condition $|\beta (U_{i\pm 1} -U_i)| \ll 1$
\cite{remark1}.
Furthermore, the condition $U''(x) \Delta x \ll 2 U'(x)$ must be fulfilled, in order to ensure the smoothness of the
potential.
In the limit  $\Delta x \to 0$ this so constructed, limiting  CTRW is described by the
fractional Fokker-Planck equation (\ref{FFPERL}), or
equivalently through Eq.~(\ref{FFPE}) \cite{letter}.


In the case of a confining potential it is sufficient to compute
the splitting probabilities $q_i ^{\pm}$ and the time
scale parameters $\nu _i$ only once over a finite $x$-region at the beginning of the
simulation. In the case of a periodic or washboard
potential, the quantities $q_i ^{\pm}$ and $\nu _i$ can be computed
only for the first period. In the latter case, while the total potential $U(x)$ is not periodic, the
potential differences appearing in the fractional rates
(\ref{f-rate}) can be rewritten as
$$
U(x_i \pm \Delta x) - U(x_i) = V(x_i \pm \Delta x) -V(x_i) \mp F \Delta x \, ,
$$
and are therefore periodic functions of $x_i$.


To perform the numerical measurements and compute the average
$\langle Y(t) \rangle$ of a quantity $Y(t) = Y(x(t))$, we
introduce a time lattice $\{ t_m^* = m \Delta t^* \}$, where $m =
0, 1, \ldots, M$, and $\Delta t^*$ is a constant time interval
between two consecutive measurements. For the computation of the
average $\langle Y(t) \rangle$, there are at least two different
strategies, which we discuss here. Both methods can be
illustrated through Fig.~\ref{algorithm}.

The first possibility is as follows: Each trajectory $x^{(n)}(t)$ is separately evolved with time, 
until the final time $t_\mathrm{final}$ is reached, $t^{(n)} \geq t_\mathrm{final}$. 
As this $n$-th trajectory reaches a measurement time $t_m^*$ 
(represented with dashed lines in Fig.~\ref{algorithm}), i.e., $t^{(n)} \geq t_m^*$, the quantity
$Y^{(n)}_m = Y(x^{(n)}(t_m^*))$ will be computed using the coordinates
corresponding to the events marked with full-circles in Fig.~\ref{algorithm}.
The value $Y_m^{(n)}$ will be saved in a storage variable
$Y_\mathrm{sum}(t_m^*) = \sum_n Y^{(n)}_m$. After evolving all the $N$ trajectories, 
the average is finally computed by normalization, 
$\langle Y(t_m^*) \rangle = Y_\mathrm{sum}(t_m^*) / N $.

The second possibility is to evolve the whole ensemble  until
the times of all the trajectories $t^{(n)}$ (at which the particles will perform the next
jump) exceed the fixed chosen measurement time $t_m^*$. We mark
these events in Fig.~\ref{algorithm} with full-circles. Then, all the
corresponding positions $x^{(n)}$ and times $t^{(n)}$ will be
saved and the average $\langle Y \rangle$ at the fixed time $t_m^*$ will be
computed. The procedure is reiterated to evolve the system until
the final time $t_\mathrm{final}$.

Which of the two methods is to be preferred depends on the problem studied
and the available computational resources.
We use the method in which the whole ensemble is evolved
in time, since it allows one to save the system configuration (and
therefore to stop and also restart the time evolution) and compute the
average quantities after each measurement time ${t_{m}^{*}}$. Furthermore, evolving
the whole system together allows one to simulate a set of $N$ particles
interacting with each other. This method also allows for
single-trajectory averages.


\subsection{ Mittag-Leffler \emph{versus} Pareto } \label{sec-MLP}

According to the Tauberian theorems \cite{Weiss}, for every $0<
\alpha <1$ the long time behavior of the system is determined
solely by the tail of the residence time distribution
\cite{MainardiFNL}. Therefore, any other distribution with the
same asymptotic form $S_\alpha(\nu_i\tau) \sim
1/\Gamma(1-\alpha)(\nu_i\tau)^{\alpha}$ could be used in place of
the Mittag-Leffler distribution (\ref{MLsurv}). In fact, also the
conditions $S_\alpha(0) = 1$ and $S_\alpha(x \to \infty) = 0$ must be satisfied, and the
function $S_\alpha(\nu_i\tau)$ has to decrease monotonically with $\tau$.

The Mittag-Leffler function $E_\alpha(-\xi)$, defined by
Eq.~(\ref{MLsurv}), can be numerically computed at $\xi < \xi _0$ through the sum
\begin{equation}\label{MLx}
E_\alpha (-\xi) \approx \sum_{h=0}^H \frac{(-\xi)^h}{\Gamma(1 +
\alpha h)} \, ,
\end{equation}
while at values of $\xi > \xi _0$ its asymptotic expansion can be used,
\begin{equation} \label{MLX}
E_\alpha (-\xi) \approx - \sum_{k=1}^K \frac{(\xi)^{-k}}{\Gamma(1
- \alpha k)} \, ,
\end{equation}
with suitable values of $H$, $K$, and $\xi _0$.

A suitable choice for an approximate description is  a Pareto probability density, defined by
\begin{equation} \label{pareto}
\psi_i(\tau) = -\frac{\mathrm{d}}{\mathrm{d}\tau} P_\alpha (\nu_i
\tau) \, ,
\end{equation}
with the survival probability
\begin{eqnarray} \label{Psurv}
P_\alpha (\nu_i \tau) &=&
\frac{1}{\left[1+\Gamma(1-\alpha)^{1/\alpha}\nu_i
\tau\right]^\alpha } \, .
\end{eqnarray}
\begin{figure}[b]
\centering
\includegraphics{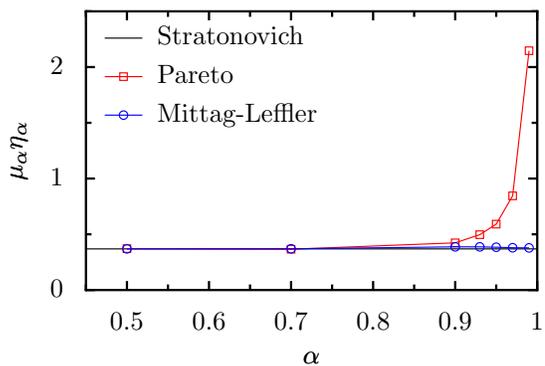}
\caption{
\label{ML2_b}
(color online).~Rescaled mobility $\mu_\alpha(F) \, \eta_\alpha$
for $F/F_\mathrm{cr}=1$, as a function of $\alpha$.
After the same rescaled simulation time $t \approx 200$, the distance from
the asymptotic limit predicted by the Stratonovich formula
(continuous line) of the mobility values corresponding the Pareto residence time
density (squares) increases significantly for $\alpha > 0.9$,
respect to those of the Mittag-Leffler residence time density (circles).
}
\end{figure}

In the simulations of the CTRW we have
usually employed the Pareto distribution $y = P_\alpha(\nu_i\tau)$.
It is convenient numerically because it can be readily inverted to
provide a random residence time $\tau$ \cite{Press1992a},
\begin{equation}
\tau = 
\nu_i^{-1}\frac{y^{-1/\alpha} - 1}{\Gamma(1 - \alpha)^{1/\alpha}} \, ;
\end{equation}
$y$ is a uniform random number in $(0,1)$. Instead, the
Mittag-Leffler distribution requires a specific algorithm to be
inverted, e.g., we have used a look-up table with the values
computed from Eqs.~(\ref{MLx}) and (\ref{MLX}).

We have numerically verified the equivalence of the Mittag-Leffler and the
Pareto distribution in the computation of the asymptotic quantities.
However, the difference in the behavior of the Mittag-Leffler and the
Pareto residence time distribution in the limit $\alpha \to 1$ has
to be noticed: Namely, for $\alpha =1$ the Mittag-Leffler
distribution transforms into the exponential function,
$E_1(-\nu_i\tau)\equiv\exp(-\nu_i\tau)$, while the Pareto
distribution remains of a power-law type, leading to normal and
anomalous diffusion, respectively. For this reason, when studying numerically
fractional diffusion with $\alpha \to 1$ the Mittag-Leffler probability distribution should be used preferably.

As the Tauberian theorems ensure the equivalence of the
Mittag-Leffler and the Pareto distributions only in the asymptotic
limit $t \to \infty$, it is to be expected that at finite
times $t$ the two choices for the probability densities provide different results. The
difference increases as the parameter $\alpha$ approaches the
value $\alpha=1$. This situation is illustrated through the example in
Fig.~\ref{ML2_b}.


\subsection{Summary of the algorithm} \label{sec:algorithm}

Here we provide the core scheme of the time evolution algorithm used in
the simulations and described above in Sec.~\ref{sec-Alg} and
Sec.~\ref{sec-MLP}. The core of the program is the following one:\\ \\ 
For every measurement time $t_m = m\Delta t^*$, where $m = 1,
\dots, M$, the loop over trajectories is performed:
\begin{itemize}
  \item For every trajectory $n$, where $n = 1, \dots, N$, the
following procedure is performed:
    \begin{itemize}
    \item[$\diamond$]  While the next jumping time is smaller than the next
measurement time,
    $t^{(n)} < m\Delta t^*$, the following steps are reiterated:
    \begin{itemize}
      \item[--]
      From Eq.~(\ref{q-mu}) the time scale parameter $\nu_i$ 
      at the current position $i$ is computed.
      A random waiting time $\tau$ is extracted from the residence time
      distribution, see Sec.~\ref{sec-MLP}, and the
      next jumping time is computed, $t^{(n)} \to t^{(n)} + \tau$.
      \item[--]
      From Eqs.~(\ref{q-nu}) the probabilities $q^\pm_i$ to perform the jump 
      from site $i$ to site $i \pm 1$ are computed.
      A uniform random number between 0 and 1 is extracted to determine
      whether the particle jumps to the right or left and the new position 
      of the particle is then computed,
      $x^{(n)} \to x^{(n)} \pm \Delta x$.
    \end{itemize}
  \item[$\diamond$] The coordinate $x^{(n)}$
  and the next jumping time $t^{(n)}$ are stored.
  \end{itemize}
  \item Statistical averages at time $t_m = m\Delta t^*$ are computed
using the stored coordinates $\{x^{(n)}\}$.
  \end{itemize}


\section{Fractional current and generalized nonlinear mobility in washboard potentials}
\label{current}

Starting out with the fractional Fokker-Planck equation in the form
(\ref{FFPE}) one can derive the expression for the mean particle
position in an one dimensional tilted periodic potential
(\ref{totpot}) \cite{letter}, reading,
\begin{eqnarray} \label{x_per}
\langle x(t) \rangle = \langle x(0) \rangle + \frac{
v_\alpha(F)}{\Gamma( \alpha + 1)} t^{\alpha} \, ,
\end{eqnarray}
where the stationary fractional current is given by,
\begin{eqnarray}\label{stratSUB}
v_\alpha(F)  = \! \! \frac{ \kappa_{\alpha} L \, [1 -
\exp(-\beta F L)]}{\int_{0}^L \mathrm{d} x \int_{x}^{x+L}
\mathrm{d} y \, \exp(-\beta[U(x) - U(y)])} \, .
\end{eqnarray}
This formula represents the anomalous counterpart of
the current known for normal diffusion and reduces to the
Stratonovich formula for $\alpha  = 1$ \cite{stratonovich,HTB}.
For completeness and for the reader's convenience a simple
derivation of Eq.~(\ref{stratSUB}) is provided in Appendix~\ref{app_j}.

Numerically, the fractional current is computed in the following manner,
\begin{eqnarray} \label{numcurr}
v_\alpha (F) = \Gamma (\alpha + 1) \lim_{t \to \infty}
\frac{\langle x(t) \rangle - \langle x(0) \rangle} { t^\alpha} \, .
\end{eqnarray}
The generalized nonlinear mobility is defined as
\begin{eqnarray} \label{nummob}
\mu_\alpha(F) = v_\alpha(F)/F \, ,
\end{eqnarray}
where $F \neq 0$.

\begin{figure}[t]
\centering
\includegraphics{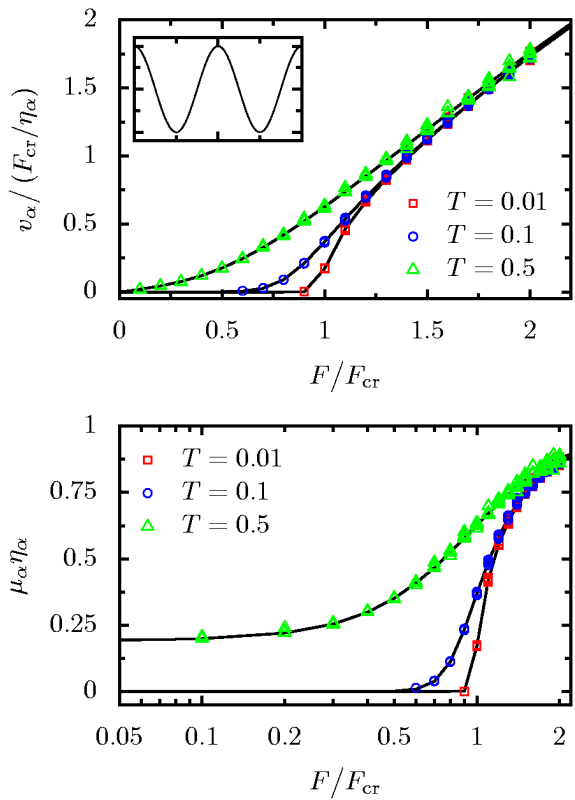}\\
\caption{ (color online).~Dimensionless subcurrent $v_\alpha (F) / (F_\mathrm{cr} / \eta_\alpha)$ and nonlinear mobility
$\mu_\alpha(F) \, \eta_\alpha$ for the case of
the cosine substrate potential (depicted in the inset) \textit{versus}
$F/F_\mathrm{cr}$. Numerical values corresponding to different
temperatures $T$ and fractional exponents
$\alpha \in [0.1, 1]$ (symbols) fit the analytic predictions from
Eq.~(\ref{stratSUB}) (continuous lines).
}
\label{J1}
\end{figure}

We test the validity of the generalized Stratonovich formula
(\ref{stratSUB}) obtained theoretically, through the simulation of
the fractional CTRW in 
different periodic potentials.
We start out with  (i) the symmetric  cosine potential,
\begin{equation} \label{cos}
V_1(x) = \cos(2\pi x / L) \, .
\end{equation}
As another type (ii) we consider  the symmetric double hump  periodic potential
\begin{equation} \label{dh}
V_{2} (x) = [\cos(2\pi x / L) + \cos(4\pi x / L)] / 2 \, .
\end{equation}
In order to explore  the role of symmetries of the substrate potential we also
consider (iii) the  asymmetric (i.e. no reflection symmetry holds), ratchet-like periodic potential, reading,
\begin{equation} \label{ratchet}
V_{3} (x) = [ 3 \sin(2\pi x / L) + \sin(4\pi x / L)] / 5 \;.
\end{equation}
The potentials in Eqs.~(\ref{cos}), (\ref{dh}), and (\ref{ratchet}), as well as the thermal energy $k_\mathrm{B}T$, are measured in the same energy unit.
For the sake of simplicity, the same symbol $T$ is used in the following to represent the rescaled thermal energy.

In the numerical simulations we have used the Pareto probability density~(\ref{pareto})
for $0 < \alpha \leq 0.8$ and the Mittag-Leffler
density (\ref{ML}) for $0.8 < \alpha < 1$ (see the discussion
in Sec.~\ref{sec-MLP}). For $\alpha = 1$ corresponding to a normal
Brownian process we have employed the exponential residence time
probability density
\begin{equation} \label{exp}
\psi_i(\tau) = -\frac{\mathrm{d}}{\mathrm{d} \tau}
\exp(-\nu_i\tau) \, .
\end{equation}
As a space step we used $\Delta x = 0.001$, measured in units of
the space period $L$. For the ensemble average we have employed $10^4$
trajectories, each one starting from the same initial condition
$x(t_0) = x_0$. The force is measured in units of the critical tilt
$F_\mathrm{cr}$, which corresponds to the disappearance of
potential extrema. In the case of the asymmetric ratchet
potential the positive critical tilt is used.

\begin{figure}[t]
\centering
\includegraphics{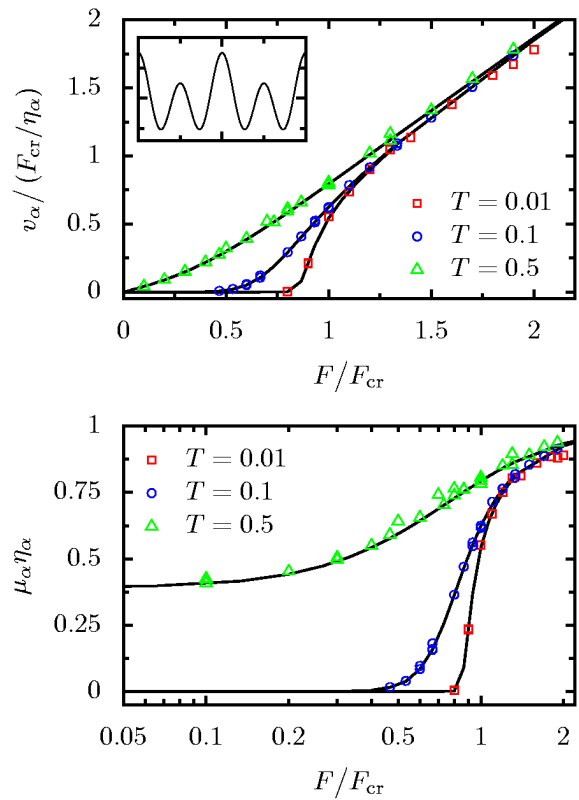} \\
\caption{ (color online).~Same as in Fig.~\ref{J1}, for the double hump potential
$V_2(x) = [\cos(x) + \cos(2x)] / 2$.}
\label{J2}
\end{figure}

\begin{figure}[t]
\centering
\includegraphics{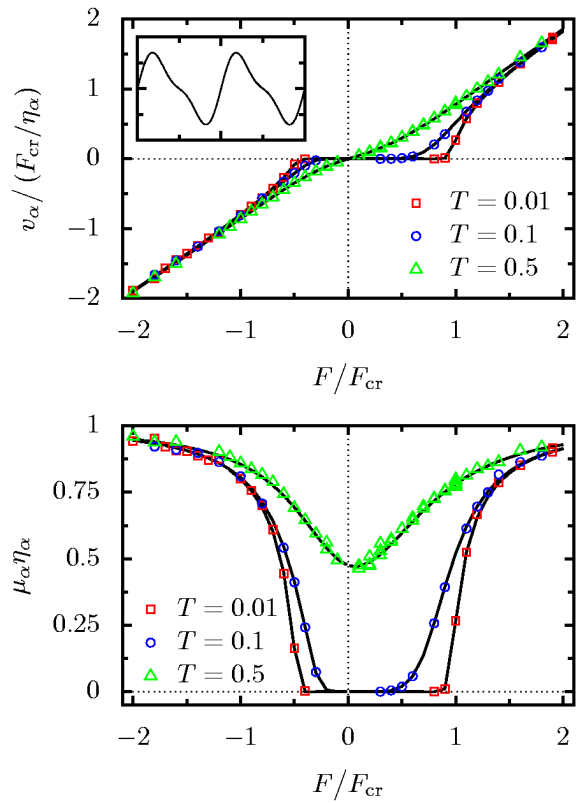}\\
\caption{
(color online).~Same as in Fig.~\ref{J1}, for the ratchet potential
$V_3(x) = [ 3 \sin(x) + \sin(2x)] / 5$.
Here also negative tilting is studied.
}
\label{J3}
\end{figure}


We present in Figs.~\ref{J1}, \ref{J2}, and \ref{J3} the numerical
results for the scaled fractional current $v_\alpha (F) / (F_\mathrm{cr} / \eta_\alpha)$
and the corresponding scaled nonlinear mobility; i.e. $ \mu_\alpha(F) \, \eta_\alpha$,
with $v_\alpha (F)$  and $\mu_\alpha(F)$ defined through Eq.~(\ref{numcurr}) and Eq.~(\ref{nummob}).
The subcurrent is measured in units of
$F_\mathrm{cr} / \eta_\alpha$, i.e. the subcurrent of a
particle under the action of a constant bias $F = F_\mathrm{cr}$,
the mobility is in units of the free mobility $\eta_\alpha^{-1}$.
Without  loss of generality we have chosen $F > 0$ for the symmetric substrate
potentials (\ref{cos}) and (\ref{dh}). In the case of the
ratchet-like potential (\ref{ratchet}) also the results
for negative values of the tilting force $F$ are depicted.

We have computed the fractional current and mobility for various
values of $\alpha$ in the interval $[0.1, 1]$. Remarkably, they do not
depend on the value of the fractional exponent
$\alpha$. For a given temperature $T$, all numerical values of
$v_\alpha (F) / (F_\mathrm{cr} / \eta_\alpha)$ and $ \mu_\alpha(F) \, \eta_\alpha$ (depicted with symbols in
Figs.~\ref{J1}, \ref{J2}, and \ref{J3}), coincide with the
theoretical curves resulting from Eq.~(\ref{stratSUB}) (continuous lines).

The regime of linear response at low temperatures is numerically
not accessible. In this parameter regime the
corresponding escape times governing the transport become far too
large \cite{HTB} and particles are effectively trapped in the
potential minima. At values of the tilting force $F$ close to
critical at which the minima disappear, the particles become
capable to escape from the potential wells and the current is
enhanced. The higher the temperature, the smaller is the tilting
required to allow the particles to escape (compare the curves corresponding
to different temperatures $T$ in Figs.~\ref{J1}, \ref{J2}, and \ref{J3}).
At higher values of the temperature $T$ the linear response regime
is numerically observable.


For tilting forces $F \gg F_{\mathrm{cr}}$ or for $T \gg 1$ the
dynamics approaches the behavior of a free CTRW that  is
exposed to a constant bias \cite{letter, scher75,
shlesinger}.


\section{ Probability densities }
\label{distribution}

\begin{figure*}[ht]
  \centering
  \includegraphics[width=6.75in]{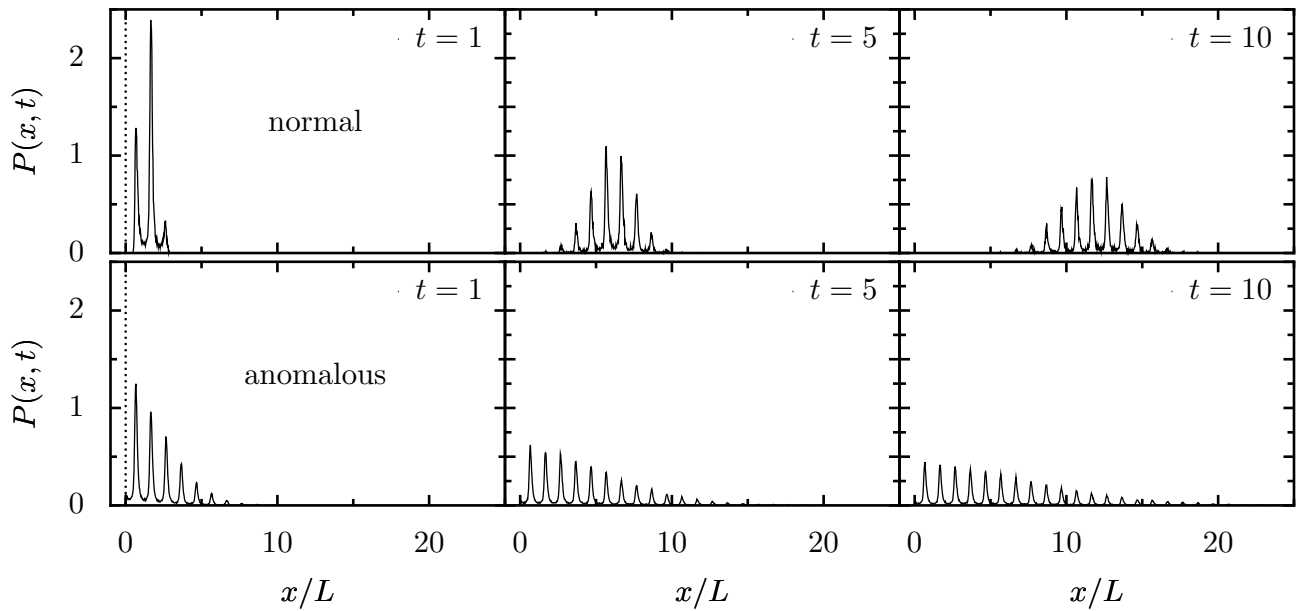}
\caption{
Comparison between the time evolution of the probability density $P(x,t)$ in the case of normal diffusion (above)
and anomalous diffusion with $\alpha = 0.5$ (below)
in a tilted periodic potential at various evolution times $t$.
This set up has been calculated for the following parameter set: $k_\mathrm{B} T = 0.1$, $F/F_\mathrm{cr} = 1$,
and a cosine potential. Dotted lines at $t=0$ represent the
initial conditions $P(x,0)=\delta(x)$; i.e., all particles start out from
the same position $x=0$.
}
\label{density}
\end{figure*}

The formal analogy between the normal and the fractional diffusion which emerges
from the similarity between the Fokker-Planck and fractional Fokker-Planck equations for the current
with the well-known Stratonovich formula  and its generalization obtained for fractional diffusion  \cite{letter}
masks some  basic physical differences. For this reason we investigate and discuss here
 the time-dependent probability density in configuration space as well as  the density of the current variable.

\subsection{ Time evolution }

The time evolution of the space probability density $P(x,t)$
in the case of anomalous diffusion is markedly different from that of normal
diffusion (see Ref.~\cite{shlesinger, scher75}).
While for normal Brownian motion, under the influence of a constant external bias,
the initial probability packet both spreads and translates on the  same time scale, one observes
in fractional diffusion mainly a spreading only towards the direction of the bias~\cite{shlesinger, scher75}.
The probability density $P(x,t)$ of particles diffusing anomalously in a washboard potential
assumes this latter feature and, at the same time, undergoes the same space-periodic modulation observed in normal
diffusion, being  typical for motion in a periodic potential.
This is illustrated in Fig.~\ref{density} in which the anomalous probability density in a washboard  potential
for $\alpha = 0.5$ (lower row) is compared with the corresponding
probability density for normal diffusion (upper row).
The data of the example in Fig.~\ref{density} have been obtained for a tilted cosine potential with $F/F_\mathrm{cr} = 1$
and at $T = 0.1$.

\subsection{ Reduced probability density} 

\begin{figure}[b]
  \centering
   \includegraphics{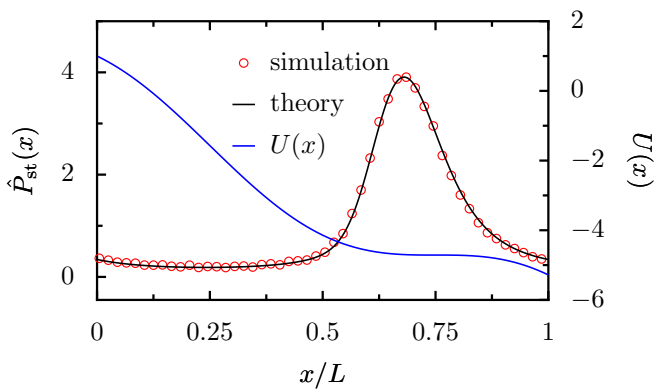}
\caption{ (color online).~Normalized theoretical stationary, reduced probability density
$\hat{P}_\mathrm{st}(x)$ for $F/F_\mathrm{cr} = 1$, $T = 0.1$, and $\alpha =
0.5$, computed from Eq.~(\ref{P1eq}) and (\ref{N}) (continuous line) and
corresponding numerical data (circles): --- left $y$-axis.  Also the underlying
potential $U(x) = \cos(x) - Fx$ is depicted: --- right
$y$-axis. }
\label{density1}
\end{figure}

The probability density $P(x,t)$ associated with a normal diffusion process in a washboard potential
cannot relax towards a stationary, asymptotic density, due to the open-boundary nature of the system.
However, the reduced asymptotic space probability density,
\begin{equation} \label{P1}
\hat{P}(x, t) = \sum_n P(nL+x, t) \, ,~n \in \mathbb{Z}  \, ,
\end{equation}
a periodic function by definition, does relax to an
asymptotic stationary density. Remarkably, in fractional diffusion, the
probability density reaches the same stationary
density as in the case of normal diffusion. The corresponding proof
follows along the same lines of reasoning leading to the asymptotic fractional current
$v_\alpha(F)$ which is formally equivalent to the Stratonovich formula
valid in normal diffusion \cite{letter}, as detailed also in the Appendix \ref{app_j}.
This result is depicted in Fig.~\ref{density1}, for the case of diffusion taking place in
a tilted cosine potential.

The form of the asymptotic reduced probability density $\hat{P}_\mathrm{st}(x)$
is given by
\begin{equation} \label{P1eq}
  \hat{P}_\mathrm{st}(x) = \mathcal{N}^{-1} \exp[-\beta U(x)] \int_x^{x+L} \mathrm{d}x' \, \exp[\beta U(x')] \, ,
\end{equation}
where $ \mathcal{N}$ is a normalization factor (see Appendix \ref{app_j}).

\begin{figure}[t]
  \centering
  \includegraphics{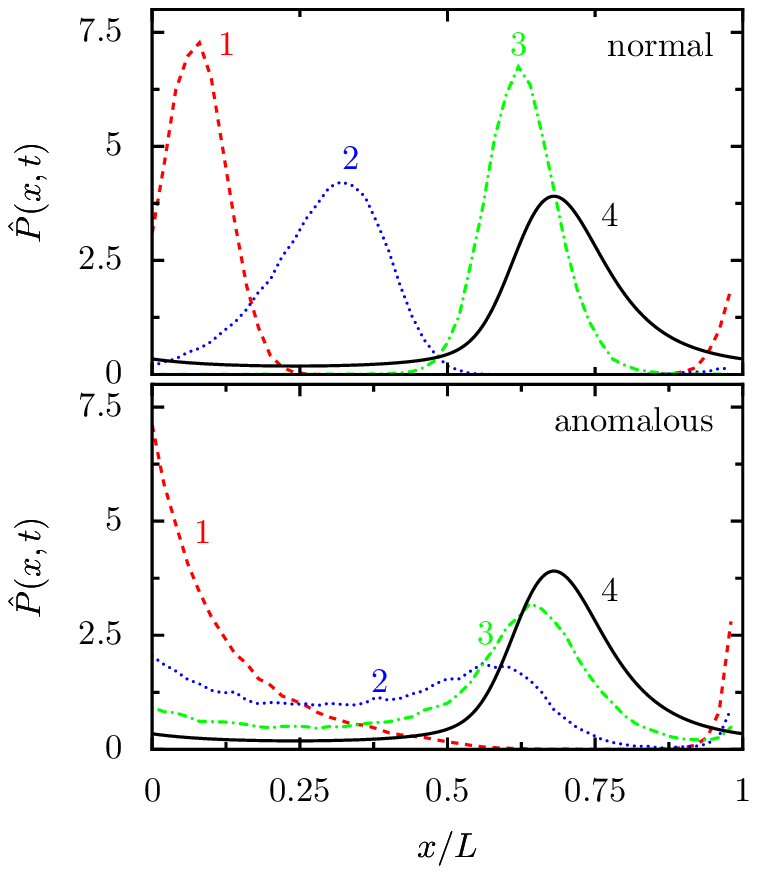}
  \caption{(color online) The time evolution for normal (above) and anomalous
(below) diffusion of the reduced probability density $\hat{P}(x, t)$ within the first
period $x \in [0,L)$ defined according to Eq.~(\ref{P1}).
In this example, the potential is $U(x) = \cos(x) - Fx$, with $F/F_\mathrm{cr} = 1$,
the temperature is $T = 0.1$, and the anomalous diffusion process corresponds to $\alpha=0.5$.
The curve labels 1, 2, 3, and 4, correspond to increasing values of time;
the solid line (theory) represents the stationary solution defined
 by Eq.~(\ref{P1eq}). }
\label{PLt}
\end{figure}

Even if the stationary probability density (depicted with continuous lines in Figs.~\ref{density1} and
\ref{PLt}) is the same, the relaxation to this stationary density is, however,   very distinct for normal and
anomalous diffusion, respectively, as shown in Fig.~\ref{PLt}.

In the case of normal diffusion, at any time instant $t$, the density
has only one maximum, which moves from the initial position ($x=0$)
toward its asymptotic position $x=x'$. At the same time it  undergoes a spreading process towards the stationary density.
As more particles reach the area around $x = x'$, the peak begins to grow,
eventually spreading again to relax to the stationary solution $\hat{P}_\mathrm{st}(x)$
(Fig.~\ref{PLt} above).

In clear contrast, for a case with anomalous diffusion the initial probability density undergoes
a spreading in the direction of the bias.
While the initial maximum of the density remains at $x=0$, a second maximum emerges
at $x \approx x'$,  which continues to grow in weight  as the density approaches the stationary shape $\hat{P}_\mathrm{st}(x)$
(Fig.~\ref{PLt} below).


\subsection{ Velocity probability density }

\begin{figure}[t]
  \centering
   \includegraphics{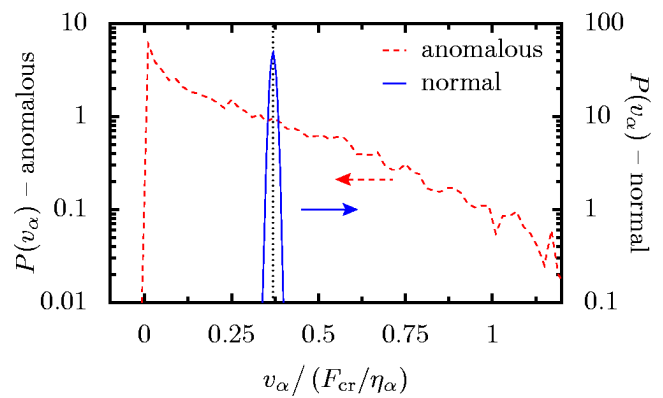}
   \caption{(color online) Anomalous (dashed line) and
     normal (continuous line)
     probability densities of the (sub)velocity, computed according to Eq.~(\ref{eq:singletrajectcurr})
     at $t=1000$ in rescaled time units. The potential, tilt, 
     temperature and $\alpha$ value are the same as in
     Fig.~\ref{PLt}. The arrows point to the
     corresponding $y$-axes.  
     Despite the very different shapes, the two probability densities
     possess the same average value as given by the (fractional) Stratonovich formula (\ref{stratSUB}),
     indicated with the vertical dotted line. 
   }
\label{fv1}
\end{figure}

For a particular trajectory realization  $x^{(n)}(t)$, the corresponding
(sub)velocity reads:

\begin{align}
  \label{eq:singletrajectcurr}
  v^{(n)}_{\alpha}:=\Gamma(\alpha + 1)
  \left[x^{(n)}(t)-x^{(n)}_{0}\right]\big/ t^{\alpha}\,  ,
\end{align}

\noindent where $n \in {1,...,N}$ and $x^{(n)}_{0}=x^{(n)}(t_{0})$. 
This (sub)velocity is a random variable and one can study the corresponding probability density.
One observes a spreading of the velocities corresponding to the broad spreading in space discussed above.
The probability density for the velocity is depicted in Fig.~\ref{fv1} for a periodic substrate cosine potential, for
$F/F_\mathrm{cr} = 1$, $T = 0.1$, and $\alpha=0.5$.
While the probability density for this velocity variable for normal diffusion (note the continuous line, right $y$-axis)
possesses a Gaussian shape, in the anomalous case this probability density  (see dashed line, left $y$-axis)
assumes a very broad  shape which falls off exponentially.
Notably, however, the two densities
have the same average, given by the Stratonovich formula,
 and indicated by the vertical dotted line in Fig.~\ref{fv1}.

\section{ Discussion and resume}
\label{conclusion}

In the field of anomalous transport the main attention thus far  has
focused on the motion under the action of a constant or
linear external force.
With this work, we have investigated the continuous time random walk with
power-law distributed residence times under the combined
action of a  space-periodic force and an external bias, thereby
elucidating in more detail parts of our previous findings
in Ref.~\cite{letter}.

The numerical algorithm for the simulation of fractional
Fokker-Planck dynamics has been detailed via the underlying CTRW. 
The application of this algorithm deserves to be commented on in greater  detail.
First, the effect of the replacement of the  Mittag-Leffler by
the Pareto distribution does not affect the
anomalous transport properties in the asymptotic limit.
However, given the finite time available for doing  simulations a difference can still
be present if the  parameter $\alpha$ assumes values close to one, i.e. close to the limit of normal diffusion.
Here, the use of the Mittag-Leffler distribution, that precisely
matches the fractional Fokker-Planck description,  is used preferably.
Otherwise, one must increase the overall time of simulations to arrive
at convergent  results.
In order to study the fractional  diffusion problem on the whole time scale, the use of
the Mittag-Leffler probability density is thus unavoidable.\\

Secondly, the weak ergodicity breaking \cite{Bel2005a} makes it impossible to
obtain the averaged value of anomalous current with a single time-average
over a single particle trajectory. Here occurs a profound difference
with the case of normal diffusion. Such a  time-averaged quantity is
itself randomly distributed, as shown by the broad density in Fig.~\ref{fv1}.
In clear contrast to the situation with normal diffusion, for anomalous diffusion the 
current probability density is very broad  and with a peak at the zero. Nevertheless,
the average value of the current agrees very well
with the theoretical Stratonovich value, as given by
Eq.~(\ref{stratSUB}). These results in turn are close in spirit to  recent work
by Bel and Barkai on the weak ergodicity breaking for a
spatially confined fractional  diffusion \cite{Bel2005a}.

The results on the averaged current and its density
are complemented by the those concerning the spatial
probability density of the particles ensemble as described
by the fractional Fokker-Planck equation,
both in the time-dependent case and for the appropriately chosen stationary regime.
The strongly asymmetric character of the spreading process of the probability densities
in Fig.~\ref{density} for the anomalous  situation is clearly in line with the nonergodic character
of the transport discussed above.

Furthermore, a new intriguing challenge concerns the universality
class of the single-particle current probability densities
within the CTRW underlying the fractional Fokker-Planck description.
The ensemble averaging is indispensable for obtaining a mean value of
the anomalous current in agreement with the theoretical results, as
shown for various potential shapes.
In particular the results depicted in Fig.~\ref{J3} imply the emergence
of rectification properties when the potential tilt is alternating in time.
The anomalous diffusion ratchet problem is, however, highly nontrivial and complex
because of intrinsic aging effects. For this reason, this objective is left for a future,
separate study. We remark that it is presently not obvious whether an adiabatic driving limit exists
at~all.

We are confident that future work will help to clarify further and  shed more light onto all these
intriguing  issues and problems.

\acknowledgments
We thank E. Barkai for useful discussions. This work has been supported by the ESF
STOCHDYN project and the Estonian Science Foundation through grant
no. 5662 (EH), the DFG via research center, SFB-486, project A10,
the Volks\-wagen Foundation, via project no. I/80424.


\appendix

\section{ Stationary (sub)current and probability density in a washboard potential }
\label{app_j}

\noindent We start from the reduced probability density and the corresponding current, i.e.,

\begin{eqnarray} \label{A1}
\hat{P}(x, t) &=& \sum_n P(nL+x, t) \, ,
\\
\hat{J}(x, t) &=& \sum_n J(nL+x, t) \, ,~n \in \mathbb{Z}  \, .
\end{eqnarray}

By definition these functions obeys periodic boundary conditions,
$\hat{P}(x+L, t) = \hat{P}(x, t)$ and $\hat{J}(x+L, t) = \hat{J}(x, t)$.
If $P(x, t)$ is normalized, e.g. $\int_{-\infty}^{+\infty} \mathrm{d}x \, P(x, t) = 1$, then
$\hat{P}(x, t)$ preserves the same normalization in any $x$-interval $(x_0, x_0 + L)$.
The condition of stationarity, obtained by letting the Caputo derivative equal to zero
in the continuity equation,
\begin{eqnarray} \label{A2}
D_{*}^{\alpha}  \hat{P}(x, t) =  - \frac{\partial \hat{J}(x, t)}{\partial x}  \, ,
\end{eqnarray}
defines the reduced equilibrium probability density $\hat{P}_\mathrm{st}(x)$.
For both normal and fractional diffusion equations,
by integrating the resulting expression in $x$,
one obtains $v_\alpha / L = \hat{J}_\mathrm{st}(x)$, i.e., explicitly,
\begin{equation}
  -\frac{v_\alpha}{L} = \kappa_\alpha \exp\left[ -\beta U(x) \right]
  \frac{d}{dx}
  \left\{
  \exp\left[ \beta U(x) \right] \hat{P}_\mathrm{st}(x)
  \right\}
  \, ,
\end{equation}
where the integration constant $v_\alpha / L$ is the one-dimensional flux.
Multiplying both sides by $\exp\left[ \beta U(x)
\right]$, integrating again  between $x$ and $x+L$, and using the
conditions $\hat{P}_\mathrm{st}(x+L) = \hat{P}_\mathrm{st}(x)$ and $U(x+L) =
U(x) - \beta F L$,
\begin{eqnarray}\label{A3}
  &&-v_\alpha L^{-1} \int_x^{x+L} dx' \, \exp\left[ \beta U(x') \right] \nonumber \\
  &&= \kappa_\alpha \exp\left[ \beta U(x) \right] \hat{P}_\mathrm{st}(x)
  \left[ \exp(-\beta F L) - 1 \right]
  \, .
\end{eqnarray}
One can now multiply both sides by $\exp\left[ -\beta U(x)
\right]$ and perform the  final integration between $x=0$ to $x=L$.
Using the normalization of $\hat{P}_\mathrm{st}(x)$ in the $x$-interval $(0,L)$,
\begin{eqnarray}
  &&-v_\alpha L^{-1}
  \int_0^{L} \mathrm{d}x \, \exp\left[ -\beta U(x) \right]
  \int_x^{x+L} \mathrm{d}x' \, \exp\left[ \beta U(x') \right] \nonumber \\
  &&= \kappa_\alpha \left[ \exp(-\beta F L) - 1 \right]
  \, .
\end{eqnarray}
From here one can obtain the explicit expression for the current,
\begin{eqnarray}
  v_\alpha =
  \frac
  {\kappa_\alpha L \left[ 1 - \exp(-\beta F L) \right]}
  {\int_0^{L} \mathrm{d}x \, \exp\left[ -\beta U(x) \right] \int_x^{x+L} \mathrm{d}x' \, \exp\left[ \beta U(x') \right]}
  \, .\\ \nonumber
\end{eqnarray}

The stationary reduced probability density can be obtained by inverting Eq.~(\ref{A3}),
\begin{eqnarray}
  \hat{P}_\mathrm{st}(x) =  \mathcal{N}^{-1} \exp[-\beta U(x)] \int_x^{x+L} \!\! \mathrm{d}x' \, \exp[\beta U(x')] \, ,
\end{eqnarray}
where the normalization constant is given by
\begin{eqnarray} \label{N}
 \mathcal{N} = \kappa_\alpha L \left[ 1 - \exp(-\beta F L)\right] / v_\alpha   \, .
\end{eqnarray}




\begin{thebibliography}{00}
\bibitem{remark1} Under this condition one can also safely replace
$\beta (U_{i\pm 1/2} -U_i)$ with $\beta(U_{i\pm1}-U_i)/2$ without
a violation of the detailed balance condition, as done for the
discretisation procedure used in Ref.~\cite{letter}.
\end{thebibliography}
\end{document}